\documentclass[aps,
prd,twocolumn,
prabib,
showpacs,nofootinbib,showkeys,10pt]{revtex4-2}
\usepackage[table]{xcolor}
\usepackage{graphicx} \usepackage{amsmath} \usepackage{amssymb}
\usepackage{amsfonts} \usepackage{bm}
\usepackage{array}
\usepackage{siunitx}
\usepackage[singlelinecheck=false]{caption}

\usepackage{ytableau}
\usepackage{multirow}
\usepackage{mathtools}
\usepackage{tikz}

\DeclarePairedDelimiter\floor{\lfloor}{\rfloor}

\begin{document}

\newcommand{\be}{\begin{equation}} \newcommand{\ee}{\end{equation}}
\newcommand{\bea}{\begin{eqnarray}}\newcommand{\eea}{\end{eqnarray}}

\title{Quantum hash function using  discrete-time quantum walk on Hanoi network}

\author{Pulak Ranjan Giri} \email{pu-giri@kddi-research.jp}

\affiliation{KDDI Research,  Inc.,  Fujimino-shi, Saitama, Japan}

%%%%%%%%%%%%%%%%%%%%%%%%%
\begin{abstract} 
Quantum walk based hash functions  have  attracted a  lot of attention in recent years because of its  faster execution time  and robust resistance against attacks  compared to classical hash functions.  It has been observed that the underlying graph and the way message controls   the quantum walk iteration steps  play a crucial role for the robustness of the hash function.  We propose a quantum hash function  based on   the discrete-time quantum walk   on  a  Hanoi network--a one  dimensional periodic lattice with extra long-range edges of a specific form--which is highly collision resistant.   The message bits  of  our scheme  control  the  flow of probability amplitude through the extra long-range edges and  the  conditional shift operators.   Our method  even works for messages with  small bit-lengths, contrary to most of the  quantum walk based hash functions defined on a cycle, which usually   work for messages with  bit-lengths  more than the  length of the cycle. 
\end{abstract}

%\pacs{03.67.Ac, 03.67.Lx, 03.65.-w}

\date{\today}

\maketitle 

%\newpage

%\tableofcontents
%%%%%%%%%%%%%%%%%%%%%%%%%
\section{Introduction} \label{in}
%%%%%%%%%%%%%%%%%%%%%%%%%
Hash function  \cite{kunth}--a mathematical  function--maps  an  arbitrary length  message, $m$,  to  a fixed length  number   $h(m)$--known as the hash value.   The process of generating hash function is  such that, given a message, $m$, it is  easy to obtain its corresponding  hash value,  $h(m)$, but from  a hash value it  is  very difficult to recover  the original   message.   Hash function has several important applications, including message authentication,  public key infrastructure,  and  digital signature.  There exist several classical hash functions, such as  SHA-1, SHA-256, SHA-512,  and MD5. However, some of them   are vulnerable to attacks and suffer from  security issues,  which  lead to  the search  beyond  the  realm  of classical  algorithm. 

Quantum computing \cite{ni}, on the other hand,  offers  a  great opportunity to develop  algorithms, which are fast and secure.  For example, Grover's algorithm \cite{grover1,grover2,giriqip} to search  unsorted database quadratically faster  than classical algorithm and Shor's  \cite{shor1,shor2}  prime number factorization  algorithm running in polynomial time are  some of the well known quantum algorithms to mention.  Quantum hash functions \cite{buh,gav,abl,abl1,zia,vas,hata} offer  unparalleled   security based on the properties  of quantum mechanics such as  superposition, and  entanglement.  The basic element of  a  quantum algorithm is the initial  quantum state, which is an  element of the Hilbert space.  In the case of quantum hash function,  quantum states, obtained from different messages,  need to be well separated in oder to avoid collision.  It also obey other properties of the hash functions, which include  easy  to obtain hash value, deterministic nature of the  generated  hash  values,  hard to obtain message from the  hash value etc.    

Quantum walk \cite{portugal}--a universal quantum  computing primitive--is   a  suitable option for  many quantum algorithms, such as quantum walk search \cite{amba2,childs, wong} with some of the  latest works reported in refs. \cite{giriijtp,giriepjd,giriijqi}, element distinctness \cite{amba} and many more  including  edge detection of digital images \cite{giripla}.  

Recently,  discrete-time quantum walk   based schemes for the hash function have attracted a  lot of attention \cite{li,li1,yang,shr,dan,hou,cao,yang1,dli,zhou1,yangijtp}.  Mostly a one-dimensional periodic lattice is involved for these   schemes.  For example,  in ref.  \cite{li},  quantum walk by   two interacting particles on  a one-dimensional periodic lattice,  controlled by the message,    has been exploited to  obtain  hash function.  The coin operator  is controlled by the message bits, i.e.,   depending on whether the   bit value  is  $0$  or $1$,    the coin operator of the corresponding iteration  involves $I$ interaction or $\pi$-phase  interaction respectively.   In this work, the  authors considered   two $2\times 2$ coin operators  for the  two  coin spaces  of  the  system. However, instead of two separate coin operators of the  previous example,  in refs. \cite{li1,yang} a  $4\times 4$ coin operator is used for both the coin spaces, while the  quantum walk is performed on a one-dimensional lattice.  Also quantum hash function based on continuous-time quantum walk has been studied  \cite{wei,wei1}. 
So far, quantum walk based hash function has  been studied in diverse underlying graphs,  which include  one-dimension lattice \cite{shr}, two-dimensional   periodic lattice with alternate quantum  walk \cite{dan,hou} and Johnson graph of $J(n,1)$ \cite{cao}, which is a complete graph.  

While implementing hash function using quantum walk,  there are many aspects which can  influence the property  of  the function, which include, the graph on which quantum walk is performed, initial state of the walk,  the way  message bits control  the evolution operator, and post processing of measured  final quantum  state.   Usually,  only   the coin operator of  the evolution operator  is controlled by the message bits.  

In the present article,  we  propose a quantum walk  based hash function, where the one-dimensional periodic lattice is attached  with a special  form of long-range edges.  Specifically, we use the Hanoi network of degree four (HN4) \cite{boe1}. 
Note that,  both  the Hanoi networks   of degree three (HN3) and four (HN4) have been used to study  quantum walk search in the literature \cite{boe2,girikor}. 
Because of these long-range edges,  we can   efficiently obtain hash values for a message of small length, whose number of bits is smaller than the number of vertices of the graph for the quantum walk.  On the other hand,  most of the quantum walk based hash functions with a  one-dimensional periodic lattice work only for the  messages with bit-length  more than the number of vertices in the graph.  Also,  both the coin  and the shift operators in our method   are controlled by the message bits. It  allows us to get a hash function\textemdash  highly resistant to collision.

This  article  is organized  in the following fashion:   A    brief  description of   the  quantum walk  on HN4   network   is  presented  in section    \ref{hanoi}.  In section   \ref{hash_hanoi},  a method for  generating  the  hash function using   this  network  is presented.  In section \ref{hash}  statistical  performance  of our hash function, described in the previous section,   has  been analyzed.   Under this section,    sensitivity of the  hash values to small changes  in the messages,   diffusion and confusion property,  uniform distribution property, collision analysis and  resistance to birthday attack are discussed.  In section \ref{comp} a comparison of our quantum hash function with other quantum hash functions is  done.  Finally  we conclude in section   \ref{con} with a discussion.

%%%%%%%%%%%%%%%%%%%%%%%%%
\section{Quantum walk on Hanoi network} \label{hanoi}
%%%%%%%%%%%%%%%%%%%%%%%%%
Hanoi network (HN)  is    generated from the    sequence of   numbered disks  of  the  Tower-of-Hanoi problem \cite{toh}.  Based on the number of edges at each vertex, there exist   two important type  of networks \cite{boe1}: Hanoi network of degree three (HN3) and of degree four (HN4).  Extensions of these  typical Hanoi networks  with even more degrees  exist,  but,  only  HN3 and HN4 networks are preferable,  since they   require  the smallest number of  qubits ($\#~\mbox{qubits} = 2$) to represent the edges  of the one-dimensional periodic lattice with extra long-range edges, which is  preferable  for the implementation in NISQ devices that have  limited and qubits and low connectivity among qubits. In this article we have considered the HN4 network because the coin operator can be easily implemented in quantum circuits since  all the four basis states of the two-qubit coin space are utilized, however we have numerically checked that the quantum hash  function using the HN3 network also has similar performance.     The main   structure  of this network is a    one-dimensional   periodic lattice   of   $N_v=  2^n$ vertices.  Like the one-dimensional lattice, it has  two  regular   edges at each vertex   $1 \leq    x_v \leq  2^n$, which are   connected to its two nearest neighbor vertices.       Additionally, it  also has  long-range  edges,  which give  the small-world  structure to the Hanoi network.  

\begin{figure}
  \centering
     \includegraphics[width=0.4\textwidth]{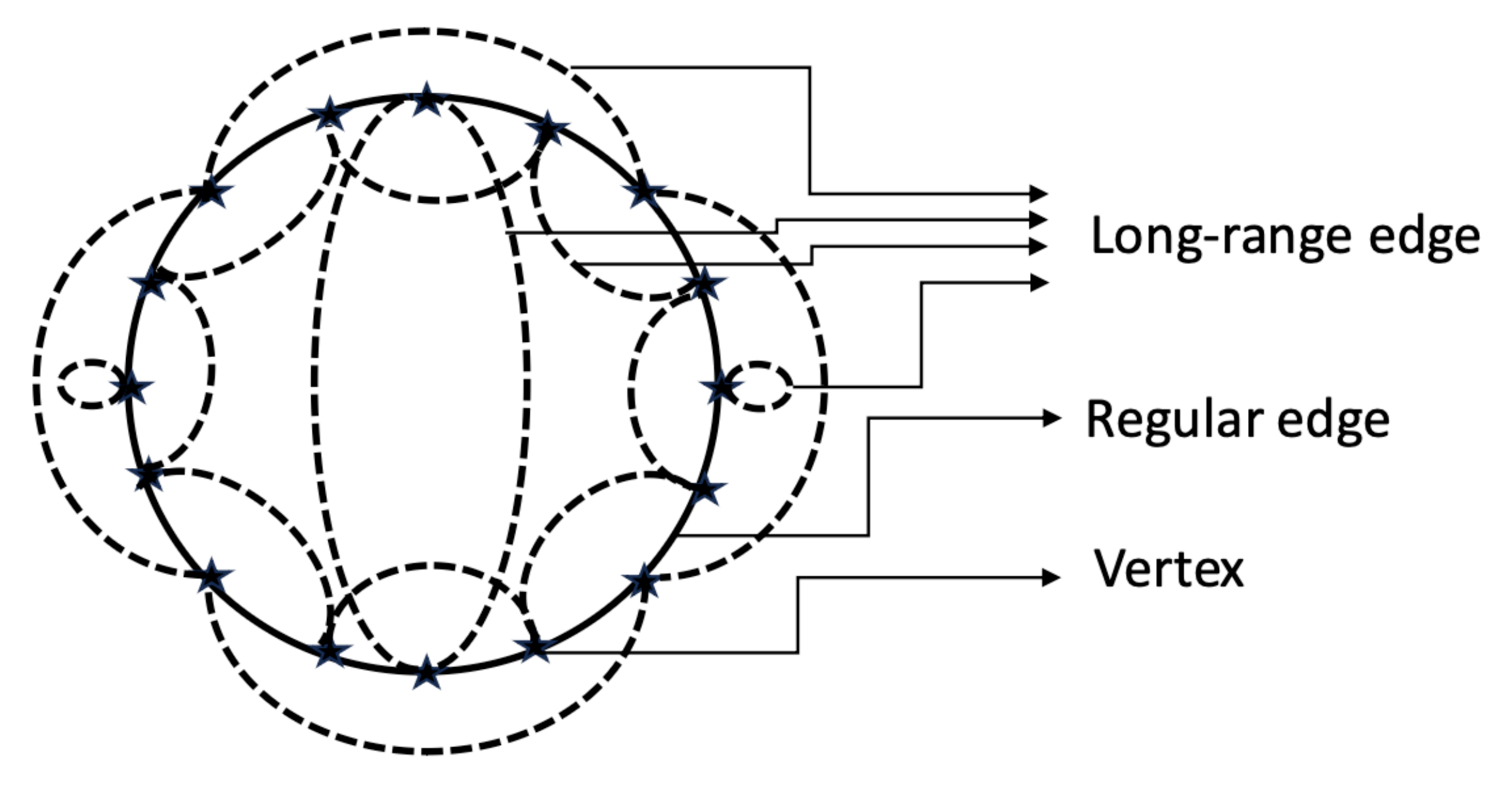}
          
       \caption{HN4 network  with $N= 2^4=  16$ vertices. Solid and dashed black  curves are regular and  long-range edges respectively.}
 \label{HN4}      
\end{figure}

Fig.  \ref{HN4} represents  a HN4 network with $N_v=16$ vertices.  The   main structure  of the network  is the  solid black   colored cycle   with sixteen vertices represented by sixteen  stars.   Long-range edges  are represented by  dashed black  color  curves.   Vertices   $x_v$ can be written  in terms of the parameters  $i$ and   $j$  as  
\begin{eqnarray}
x_v= 2^{i} \left(  2j +1  \right)\,,   
\label{hier}
\end{eqnarray}  
Note that       $0 \le i \le n$  and  $0 \le j \le  j_{max}=  \floor{2^{n-i-1} -  1/2}$ uniquely identify  the vertices   $x_v$.   Alternatively,    $N$ basis  states  $|x_v \rangle$ of the network can   be expressed as  $|x_v \rangle  =   |i,  j \rangle$.
%----------------------------------------------------------------------------------------------------
%----------------------------------------------------------------------------------------------------
There exist  four edges of the  HN4 network:   two regular  edges, which  come from the one-dimensional periodic lattice and  two extra long-range edges connecting  
$|i, j \rangle$   to   $|i, j -1  \rangle$  and  $|i, j +1\rangle$ respectively, for  $i \neq n-1, n$.  Also, there are two   self-loops  at  $|n-1, 0\rangle$ and  $|n, 0\rangle$  respectively.  In fig. \ref{HN4},    we  can  see  each vertex has  two  solid black colored  regular edges  and two  dashed black colored long-range   edges.   Two  of  the long-range   edges   form   undirected   self-loops at $|n-1=3, 0\rangle$ and  $|n=4, 0\rangle$  respectively. 
%%%%%%%%%%%%%%%%%%%%%%%%%%%%%%%%%%%%%%%%%%%%%%%%%%%%%%%%%%%%%%%%%%%%%%%%%%%%%%%%%%%%%%%%%%%%
 
The Hilbert space $\mathcal{H} = \mathcal{H}_C \times \mathcal{H}_V$ of this network is a tensor product  of the coin space  $\mathcal{H}_C$ and the vertex space  $\mathcal{H}_V$ respectively. Quantum walk  in this space   is obtained by  the coin operator, $C$, which acts   on   a   state of the $4$-dimensional coin space,  followed by the shift operation, $S$,  which acts on the combined coin and   $N_v$-dimensional vertex spaces.      
The  initial state for the quantum walk  is of the  form 
\begin{eqnarray}
|\psi_{in}\rangle =   |\psi_{c}\rangle \otimes  |\psi_{v}\rangle   \,,
\label{in}
\end{eqnarray}
where  $|\psi_{c}\rangle$ and  $|\psi_{v}\rangle$ are the normalized state of the coin and vertex space respectively. Specific form of  these states depend our choice, so that  we can generate highly secured  hash function.

The evolution operator for the quantum walk is given by  
\begin{eqnarray}
 \mathcal{U} = S\left(C \otimes I \right)\,,
\label{uqw}
\end{eqnarray}
which  acts repeatedly on the initial state  $|\psi_{in}\rangle$   to let the  state  evolve.   However, in the case of  hash function generation  we need to modify  the evolution operator  
$ \mathcal{U}$ in each iteration  step, which will be  discussed in detail in  the next section. 

\begin{figure*}
  \centering
     \includegraphics[width=0.7\textwidth]{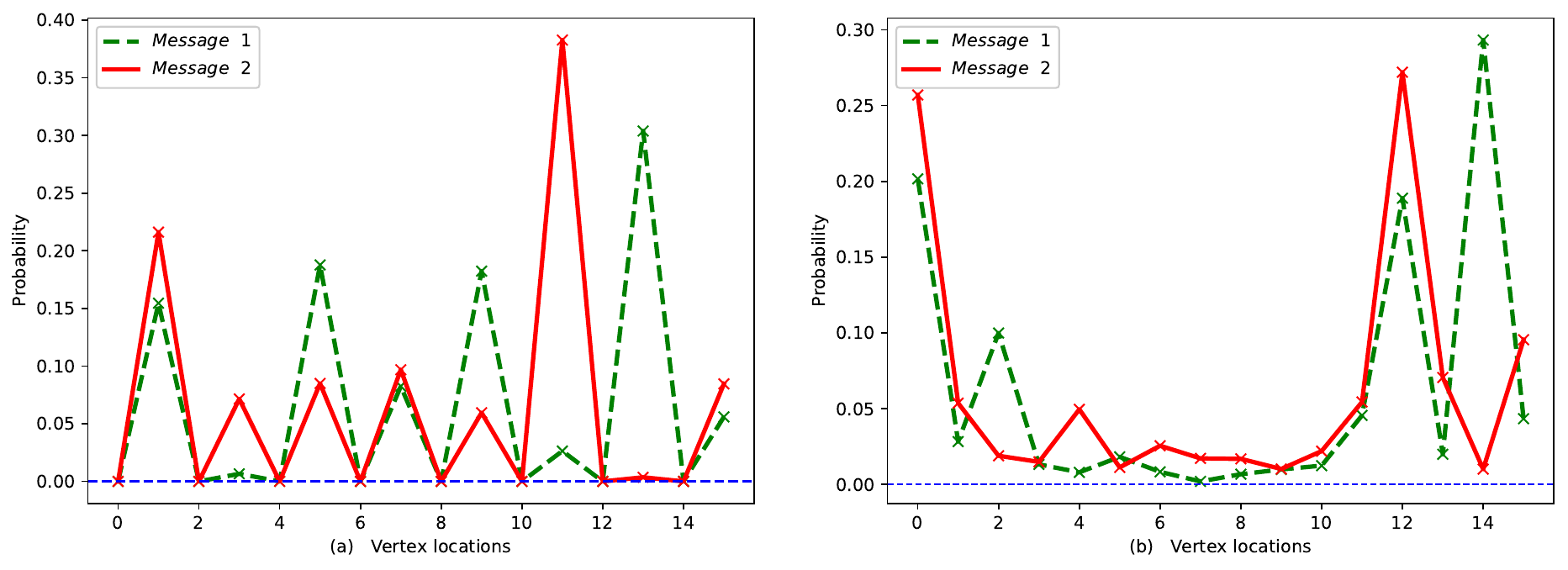}
          
       \caption{Probability distribution of a pair of randomly generated $9$-bit length messages (message $1$, message $2$)  as a function of the  vertex location for the quantum walk on  (a) a one-dimensional  periodic lattice and (b) on a HN4 network for $N= 16$, $t=9$.}
 \label{qhash_prob}      
\end{figure*}

One of the important properties of  the quantum walk on a finite one-dimensional  lattice  with $t$ time steps  is that,  for even $N$,  the probability distribution at location  $x$ is zero \cite{portugal}  when  $x+t$ is odd, as can be seen from  Fig. \ref{qhash_prob}(a). Therefore, the  probability distribution is zero at  even or odd vertices for odd or even $t$ respectively.   However, for odd $N$  the probability  distribution is nonzero at all vertex locations for large enough $t$.  So, normally, one cannot   use even-length one-dimensional periodic lattice \cite{hou}   for the  generation of a  hash function,  because it  leads to very  high  collision rates due to  the predictable periodic  zero  probability distributions.     The  HN4 network, which we use in this study,  allows us to  consider an even-length  one-dimensional periodic lattice with no periodic  zero-probability distribution as can be seen from  Fig. \ref{qhash_prob}(b).  Also, because of its two long-range  edges  we can  manipulate probability distribution through the coin operator  in such a way  that  the  generated hash function  has very  high collision  resistance.   Note that, lively quantum walk with even lively-parameter  studied in ref. \cite{hou} also  does not have   periodic  zero  probability distributions. 
%----------------------------------------------------------------------------------------------------
%----------------------------------------------------------------------------------------------------
%-------------------------------------------------------------------------------------------------------------------------------------------------------------
\begin{table*}
\centering
\begin{tabular}{ |  p{0.5cm }  | p{1.0cm } |  p{14.cm}  |}
\hline
\multirow{1}{*}{\textbf{{\scriptsize }}} & \multicolumn{2}{c|}{\textbf{ {\scriptsize  Messages/hash values}}} \\
\hline    
    
\multirow{2}{*}{$~~1$} & \cellcolor{cyan!10}  $m_1$ &  \cellcolor{cyan!10}  $111110010011000$     \\ 
\cline{2-3}
 & \cellcolor{orange!10} $h(m_1)$ & \cellcolor{orange!10}  ‘6EEA’  ‘6CA5'  ‘3C9E'  ‘7333'  ‘5A78'  ‘3598'  ‘4AFC'  ‘A19D' ‘5455' ‘6DE2'
 ‘3646'  ‘4518'  ‘4F38'  ‘7CC8'  ‘5CCF'  ‘6AAC'   \\ \hline

 \multirow{2}{*}{$~~2$} & \cellcolor{cyan!10}  $m_2$ &  \cellcolor{cyan!10}  $111110000011000$  \\ 
\cline{2-3}
 & \cellcolor{orange!10}  $h(m_2)$ & \cellcolor{orange!10}  ‘858C' ‘7492' ‘7CB8' ‘4377' ‘3CBE' ‘1F10' ‘6794' ‘1EBB' ‘8C5B' ‘46A6'
 ‘3407' ‘5F21' ‘410F' ‘8429' ‘4592' ‘937B'   \\ \hline

  \multirow{2}{*}{$~~3$} & \cellcolor{cyan!10}  $m_3$ &  \cellcolor{cyan!10}  $111110010111000$ \\ 
\cline{2-3}
 & \cellcolor{orange!10} $h(m_3)$& \cellcolor{orange!10}  ‘5D89' ‘17D0' ‘346E' ‘4933' ‘5818' ‘5D48' ‘6255' ‘53FB' ‘359F' ‘68BF'
 ‘5053' ‘73FD' ‘84A2' ‘3D01' ‘9E02' ‘942D'  \\ \hline

  \multirow{2}{*}{$~~4$} & \cellcolor{cyan!10}  $m_4$ &  \cellcolor{cyan!10}  $1111110010011000$ \\ 
\cline{2-3}
 & \cellcolor{orange!10}  $h(m_4)$ & \cellcolor{orange!10}  ‘8414' ‘9AF6' ‘423D' ‘500E' ‘5A2A' ‘49AD' ‘79AC' ‘7E39' ‘3D09' ‘39F3'
 ‘6B34' ‘5509' ‘4047' ‘61E0' ‘5511' ‘6065' \\ \hline

  \multirow{2}{*}{$~~5$} & \cellcolor{cyan!10}  $m_5$ &  \cellcolor{cyan!10}  $11111010011000$ \\ 
\cline{2-3}
 & \cellcolor{orange!10}  $h(m_5)$ & \cellcolor{orange!10}  ‘3546' ‘2B9D' ‘98E3' ‘2DDF' ‘6425' ‘3B59' ‘6351' ‘69EF' ‘87F7' ‘4B93'
 ‘A763' ‘56F7' ‘58E2' ‘17C0' ‘6ECF' ‘54F1'\\ \hline
 
\end{tabular}
\caption{Messages in binary forms and corresponding hash values in hexadecimal forms.} 
\label{table1}
 \end{table*} 
%-------------------------------------------------------------------------------------------------------------------------------------------------------------

%%%%%%%%%%%%%%%%%%%%%%%%%
\section{Hash function with quantum walk on Hanoi network} \label{hash_hanoi}
%%%%%%%%%%%%%%%%%%%%%%%%%
To exploit quantum walk for the generation of hash function,  the evolution operator   $\mathcal{U}$ needs to be  controlled by a message.  Usually,  only the coin operator  of  the evolution operator is controlled by the message bits.  However, in our approach   we    control  both the coin   and shift   operator,   based on the bit-values$(0/1)$ of a  message,  as follows:

{\it Bit-value $= 0$}:
 The  evolution operator used in this case  is the following:
\begin{eqnarray}
 \mathcal{U}_{0} = S_{0}\left(C_{0} \otimes I \right)\,,
\label{uqw}
\end{eqnarray}
where  we use the following  Grover  operator  for the  coin  
\begin{eqnarray}
C_0 =    \left(2 |\psi_0 \rangle \langle \psi_0 | - I \right)\,,
\label{gc0}
\end{eqnarray}
with 
\begin{eqnarray} \nonumber
|\psi_0 \rangle = 
\mathcal{N}_0 \left(  |0 \rangle  + |1 \rangle  + \sqrt{l_{00}} |2 \rangle  + \sqrt{\tilde{l}_{00}} |3 \rangle \right)\,,
\label{st0}
\end{eqnarray}
and  $\mathcal{N}_0 = 1/(2+ l_{00}+ \tilde{l}_{00})$. 
The  flip-flop shift operator  is composed of three parts   
\begin{eqnarray} 
 S_{0}  =   S_{00} + S_{01} + S_{02}\,.
 \label{fshift1}
\end{eqnarray}
Note that, flip-flop shift operator is essential in  spatial search on graph by discrete-time quantum walk \cite{amba2,wong}. 
The first part   of eq. (\ref{fshift1})
\begin{eqnarray} \nonumber
 S_{00} = \sum^{ N_v}_{x_v=1}  \left( |1\rangle \langle 0 | \otimes | x_v +1  \rangle \langle  x_v  |  \hspace{2.5cm}  \right.\\  \nonumber 
 \left. +  |0 \rangle \langle 1 | \otimes | x_v -1 \rangle \langle  x_v | \right)\,,   
\label{fshift11}
\end{eqnarray}
is the  usual  flip-flop shift operator associated with  the regular  edges  $|0 \rangle$ and  $|1 \rangle$  of  the   HN4 network. The second part 
\begin{eqnarray} \nonumber
 S_{01} =   \sum^{ n-2}_{i=0}   \sum^{j_{max}}_{j=0}  \left(  |3 \rangle \langle 2 | \otimes | i, j + 1  \rangle \langle  i, j |  \hspace{2cm} \right. \\ \nonumber 
\left. + |2 \rangle \langle 3 | \otimes | i, j - 1  \rangle \langle  i, j | \right)\,, 
\label{fshift12}
\end{eqnarray}
is the   flip-flop shift operator  associated with  the long-range edges  $|2 \rangle$ and  $|3 \rangle$,   and the third part 
\begin{eqnarray} \nonumber
 S_{02} =   \left( |2 \rangle \langle 2 | +   |3 \rangle \langle 3 |  \right)\otimes  \hspace{4cm}  \\ \nonumber
 \left( | n-1 , 0  \rangle \langle  n-1, 0 |  +   | n , 0  \rangle \langle  n, 0 | \right)\,,
\label{fshift13}
\end{eqnarray}
is  associated with the two self-loops   formed by the long-range edges  at $(i= n-1, 0)$, and $(i= n, 0)$ respectively. 
Note that,   the four-dimensional coin space  is formed by   two  regular edges  $|0 \rangle,  |1 \rangle$  and  two long-range edges   $|2 \rangle,  |3 \rangle $.

{\it Bit-value $= 1$}:
The  evolution operator in this case  is the following:
\begin{eqnarray}
 \mathcal{U}_{1} = S_{1}\left(C_{1} \otimes I \right)\,,
\label{uqw1}
\end{eqnarray}
where  we use  Grover  operator  of the form  
\begin{eqnarray}
C_1 =    \left(2 |\psi_1 \rangle \langle \psi_1 | - I \right)\,,
\label{gc1}
\end{eqnarray}
with 
\begin{eqnarray} \nonumber
|\psi_1 \rangle = \mathcal{N}_1
\left(  |0 \rangle  + |1 \rangle  + \sqrt{l_{11}} |2 \rangle  + \sqrt{\tilde{l}_{11}} |3 \rangle \right)\,,
\label{st1}
\end{eqnarray}
and  $\mathcal{N}_1 = 1/(2+ l_{11}+ \tilde{l}_{11})$. 
The  regular shift operator, here, like the previous case,   is composed of three parts   
\begin{eqnarray} 
 S_{1}  =   S_{10} + S_{11} + S_{12} 
 \label{fshift14}
\end{eqnarray}
The first part   
\begin{eqnarray} \nonumber
 S_{10} = \sum^{ N_v}_{x_v=1}  \left( |0\rangle \langle 0 | \otimes | x_v +1  \rangle \langle  x_v  |  \hspace{2.5cm}  \right.\\  \nonumber 
 \left. +  |1 \rangle \langle 1 | \otimes | x_v -1 \rangle \langle  x_v | \right)\,,
\label{fshift15}
\end{eqnarray}
second part 
\begin{eqnarray} \nonumber
 S_{11} =   \sum^{ n-2}_{i=0}   \sum^{j_{max}}_{j=0}  \left(  |2 \rangle \langle 2 | \otimes | i, j + 1  \rangle \langle  i, j |  \hspace{2cm} \right. \\ \nonumber 
\left. + |3 \rangle \langle 3 | \otimes | i, j - 1  \rangle \langle  i, j | \right)\,,
\label{fshift16}
\end{eqnarray}
and the third part 
\begin{eqnarray} \nonumber
 S_{12} =   \left( |2 \rangle \langle 2 | +   |3 \rangle \langle 3 |  \right)\otimes  \hspace{4cm}  \\ \nonumber
 \left( | n-1 , 0  \rangle \langle  n-1, 0 |  +   | n , 0  \rangle \langle  n, 0 | \right)\,.
\label{fshift17}
\end{eqnarray}

The final state for the hash function for a general  message $m = `b_s \cdots b_2 b_1 b_0'$ in binary form with a bit-length $s$  is of the form 
\begin{eqnarray}
|\psi_{f}\rangle =   \mathcal{U}_{b_0} \mathcal{U}_{b_1} \cdots \mathcal{U}_{b_s} |\psi_{in}\rangle \,.
\label{fi1}
\end{eqnarray}
For example,  for a message  $m= `110100'$, the final state would be   $|\psi_{m}\rangle =   \mathcal{U}_{0} \mathcal{U}_{0} \mathcal{U}_{1} \mathcal{U}_{0} \mathcal{U}_{1} \mathcal{U}_{1}|\psi_{in}\rangle$. 
The iteration steps for the generation of our hash function, obtained from  eq.  (\ref{fi1})  increases  linearly   with the   bit-length $s$ of the message.  

Although,  the results presented  in this article are  based on  noiseless analysis,  in real-life scenarios,  a  longer  bit-length, corresponding to more iteration steps,   can  transform   a  quantum probability distribution of the quantum walk into  a classical  Gaussian-like distribution   in the presence of unitary  noise \cite{dans}.  Noise associated with more iterations steps  significantly degrades the probability amplitudes in Grover search algorithms also, which can be mitigated with strategies involving  less iteration steps   \cite{kor}. 
Correction  methods such as  error correction \cite{knill}, and  decoherence free  sub-spaces \cite{kempe} require  significant overhead, which is  a luxury in  the current  NISQ era.  The effect of  noise can be minimized  by   reducing  the  number of iteration  steps of our scheme  significantly   by  adopting an approach from ref. \cite{yang1}, where each evolution operator is controlled by two bits of the message  instead of one in   eq.  (\ref{fi1}),  without  compromising  the performance of  our  hash function. 
The final  state for the hash function then becomes 
\begin{eqnarray}
|\psi_{f}\rangle =   \mathcal{U}_{b_0 b_1} \mathcal{U}_{b_2 b_3} \cdots \mathcal{U}_{b_{s-1} b_s} |\psi_{in}\rangle \,,
\label{fi1s}
\end{eqnarray}
where  $\mathcal{U}_{ij} = S_{i}\left(C_{ij} \otimes I \right)$. Four different  coin operators  $C_{00}, C_{01}, C_{10}, C_{11}$, needed for this method,  can be obtained by appropriately choosing  two parameters  associated with the long-range edges  in the  underlying  coin  state. 
For example,  for a message  $m= `110100'$, the final state would now be   $|\psi_{f}\rangle =   \mathcal{U}_{00} \mathcal{U}_{10} \mathcal{U}_{11} |\psi_{in}\rangle$. The required  iteration  steps for  the final state $|\psi_{f}\rangle$ can further be reduced by adopting  other similar methods  presented in ref. \cite{yang1}.
We acknowledge that a comprehensive performance validation under realistic noisy conditions remains an essential issue. Future work will involve detailed simulations to assess the impact of varying noise levels on collision rates, distribution uniformity, and the effectiveness of our proposed scheme. 

The final probability for the walker to be at  $x_v$  location can be evaluated as 
\begin{eqnarray}
P(x_v,m) =  \sum_{i=0}^3  |\langle i, x_v |\mathcal{U}_{b_0} \mathcal{U}_{b_1} \cdots \mathcal{U}_{b_s} |\psi_{in}\rangle|^2 \,,
\label{prob}
\end{eqnarray}
To generate the hash function, we choose an initial state  of the form
\begin{eqnarray}
|\psi_{in}\rangle = \frac{1}{2} \sum_{i=0}^3 |i\rangle \otimes |x_v=0\rangle \,,
\label{inh}
\end{eqnarray}
and  perform the following  steps:
\begin{enumerate}
\item Choose  the values of the set of  parameters  $(N_v, l_{00}, \tilde{l}_{00}, l_{11}, \tilde{l}_{11})$.  For the  experiment  performed in this paper we choose  $(16,  0.01, 1.0, 0.1, 0.01)$.  

\item Run quantum walk on Hanoi network HN4 as described  above and in the previous section. 

\item  Measure   the  the probability distribution $P(x_v,m)$ as the output.  

\item Post-processing: Multiply  probability amplitude   $\sqrt{P(x_v,m)}$ by  $10^l$.   Obtain   $h_{x_v}=\sqrt{P(x_v,m)} \times 10^l$  $mod~ 2^{k}$,  for $10^l \geq 2^k$.  The precision level, $l$,  can be increased  accordingly as the size of the network increases, which reduces the differences between probability amplitudes, requiring more decimal places  to be included.   We choose $l=5, k=16$, which makes  each $h_{x_v}$  a  $16$-bit binary string.  Concatenate all $h_{x_v}$s to obtain a $L= N_v\times 16$ bit  hash function as $h(m)= \left( h_0, h_1, \cdots, h_{N_v-1}\right)$. 
\end{enumerate}
%----------------------------------------------------------------------------------------------------
\begin{figure*}
  \centering
     \includegraphics[width=0.7\textwidth]{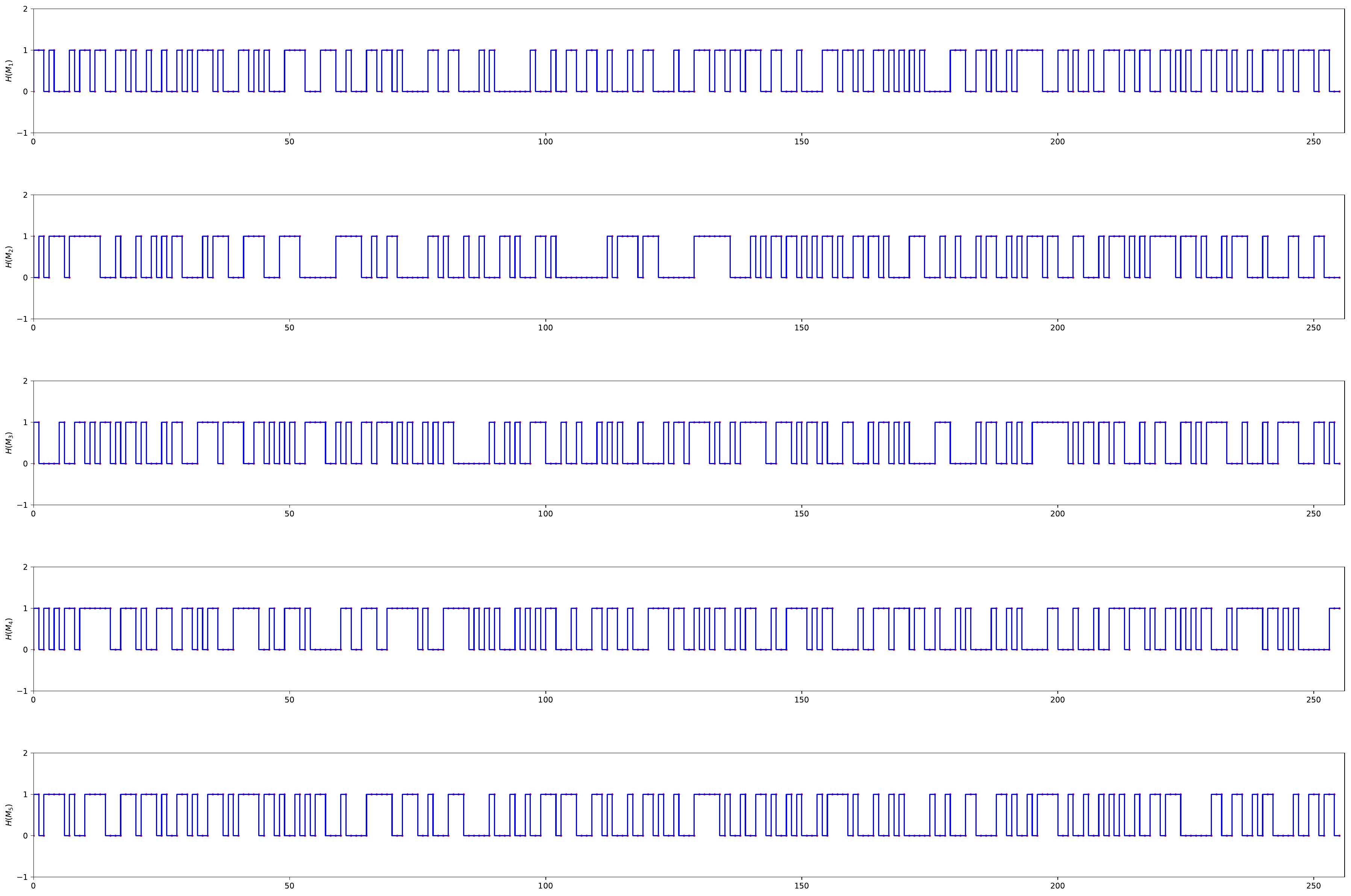}
          
       \caption{Plot of the hash values $h(m_1)$, $h(m_2)$, $h(m_3)$, $h(m_4)$,  and $h(m_5)$. Red stars are the bit-values  and blue curves are step plots of the hash values respectively.}
 \label{hash_val}      
\end{figure*}
%----------------------------------------------------------------------------------------------------

%-------------------------------------------------------------------------------------------------------------------------------------------------------------
\begingroup

\setlength{\tabcolsep}{3.0pt} % Default value: 6pt
\renewcommand{\arraystretch}{1.2} % Default value: 1
\begin{table}
 \centering

    \begin{tabular}{ |  p{1.2cm }  | p{1.5cm } |  p{1.5cm} | p{1.6cm} | p{1.5cm} |}
    \hline
      & $N=2000$ & $N=5000$  &  $N=10000$  &  \textbf{\scriptsize Mean} \\ \hline
    
    $B_{min}$ & $91$  & $95$  &  $92$  &  $92.6667$  \\ \hline
    
     $B_{max}$  &   $147$  & $148$  &  $154$ &  $149.6667$    \\ \hline
     
     $\bar{B}$ & $122.52$  & $122.6322$  &  $122.6559$  &  $122.6027$  \\ \hline
    
     $\bar{P}(\%)$  &   $47.8594$  & $47.9032$  &  $47.9125$ &  $47.8917$    \\ \hline
     
     $\Delta B$ & $8.0311$  & $8.1753$  &  $8.1115$  &  $8.1071$  \\ \hline
    
     $\Delta P(\%)$  &   $3.1372$  & $3.1935$  &  $3.1686$ &  $3.1664$    \\ \hline
    
 \end{tabular}
 \caption{Result for diffusion and confusion test.} 
 \label{table2}
 \end{table} 
 \endgroup
 %-------------------------------------------------------------------------------------------------------------------------------------------------------------

%%%%%%%%%%%%%%%%%%%%%%%%%
\section{Performance of hash  function} \label{hash}
%%%%%%%%%%%%%%%%%%%%%%%%%
In general, our proposed hash  function has  a  bit-length $L = N_v \times k$, where $N_v$ is the number of vertices $x_v$ of  the  HN4 network and $k$ is the  bit-length of each  probability amplitude   $h_{x_v}$.  There is no  conclusive analytical proof for  the security of the  hash function.  However, the process of our quantum walk based hash function involves  measurement of final quantum state in computational basis and in  the  post-processing  stage  module  $2^k$ operation,  both of which being many-to-one mappings, are  irreversible  one-way processes.  Therefore,  it is resistant to preimage attack, as given  a hash value $y$ it is hard to find its  corresponding message  $m$, such that  $H(m) = y$.  It is also resistant to second preimage attacks, since given a message $m_1$  it is computationally infeasible  because of the irreversible nature of the involved  processes, to find another message $m_1$ such that  $H(m_1) = H(m_2)$. 

In the following sub-sections we  perform  several  standard statistical  analysis \cite{li,li1,yang,shr,dan,hou,cao,yang1,dli,zhou1}, which show that  the  hash function is secured and highly collision resistant.   
%%%%%%%%%%%%%%%%%%%%%%%%%
\subsection{Sensitivity of hash values to change in message} \label{sen_hash}
%%%%%%%%%%%%%%%%%%%%%%%%%
From a random message, by  making  four different types of changes to the bit-string,   we  obtain four additional messages.  These  five messages  generate five corresponding hash values, which are  compared to show that a small change in the original message outputs   huge changes in the hash value.  The steps are the following:
\begin{enumerate}
\item Choose  a random  message, $m_1$.

\item Change a randomly chosen bit $1$  of the message $m_1$  to bit $0$  to obtain  the  message, $m_2$.

\item Change a randomly chosen bit $0$   of the message $m_1$  to bit $1$  to obtain  the   message, $m_3$.

\item Add a bit at  a random  position   of the message $m_1$   to obtain  the   message, $m_4$.

\item Delete  a bit at a  random  position   of the message $m_1$   to obtain  the   message, $m_5$.

\item Generate   hash values $h(m_1)$, $h(m_2)$, $h(m_3)$, $h(m_4)$,  and $h(m_5)$. 
\end{enumerate}
The messages and their corresponding    $256$-bit  hash values  in hexadecimal format  are presented in table  \ref{table1}.  We also plot  the hash values in fig. \ref{hash_val}, which shows that a small change in the  original message, $m_1$,  results in huge changes in the hash value,  showing   high  collision resistance property. 
%----------------------------------------------------------------------------------------------------
\begin{figure*}
  \centering
     \includegraphics[width=0.8\textwidth]{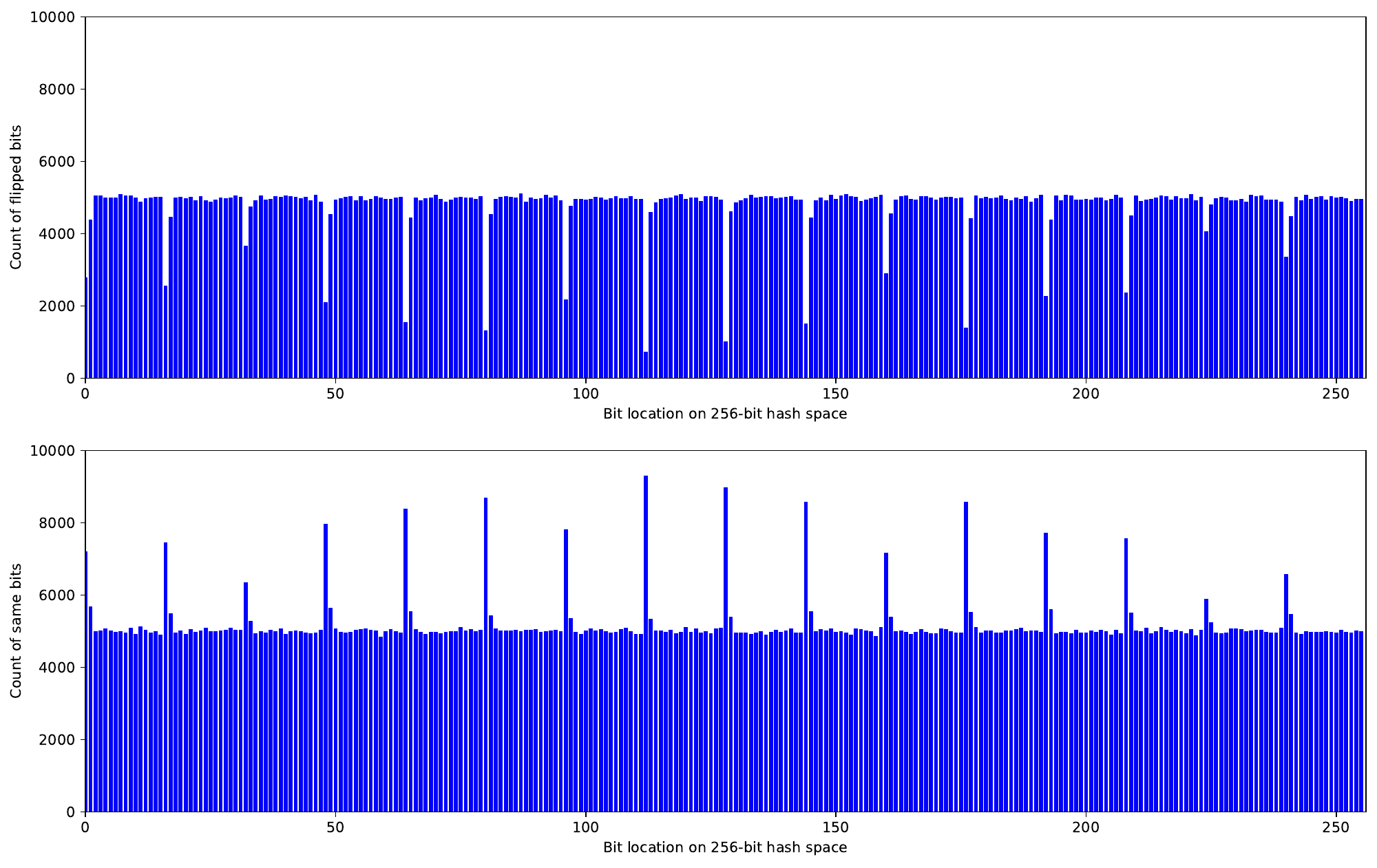}
          
       \caption{Uniform distribution of bits on  hash space}
 \label{hash_bar}      
\end{figure*}
%----------------------------------------------------------------------------------------------------

%%%%%%%%%%%%%%%%%%%%%%%%%
\subsection{Diffusion and confusion properties} \label{diff_hash}
%%%%%%%%%%%%%%%%%%%%%%%%%
In this sub-section  we study the  diffusion and confusion property  the  hash function.   The following steps are performed:
\begin{enumerate}
\item Choose  a random  message, $m_1$.

\item Change a bit at random position of the  previous message  to obtain  another  message, $m_2$.

\item Generate $256$-bit hash values, $h(m_1), h(m_2)$, in binary format,   corresponding  to the pair of messages $m_1, m_2$ respectively. 

\item Count the  total  number of flipped bit(Hamming distance)  for the two hash values.   Let  $h(m_1) = \left(b_1^1, b_2^1, \cdots, b_L^1\right)$, $h(m_2) = \left(b_1^2, b_2^2, \cdots, b_L^2\right)$ are  the hash values in binary form.  Then  Hamming distance  $B_i$,  for the  $i_{th}$  experiment among  the total  $N$ experiments is given by 
$B_i = \sum_{j=0}^{L-1} |b_j^1 -  \bar{b}_j^2 | $. 

\item Repeat  $1. - 3.$   $N$ times and calculate the following:

\item  $B_{min} = \scriptsize{min}\left(\{ B_i\}^N_1\right)$. 

\item  $B_{max} = \scriptsize{max}\left(\{ B_i\}^N_1\right)$. 

\item  $\bar{B} =  \left( \sum_{i=1}^N B_i \right)/N$. 

\item $P = \bar{B}/ L$, $P(\%) = P\times 100$

\item $\Delta B=  \sqrt{  \sum_{i=1}^{N}  (B_i -\bar{B})^2}/(N-1) $

\item $\Delta P(\%)=  \sqrt{  \sum_{i=1}^{N} (B_i/L  - P)^2/(N-1)}\times 100 $
\end{enumerate}
Diffusion and confusion test are  performed for $N=2000, 5000$ and $10000$ and  results  presented  in table \ref{table2}  are   close to the ideal values.

%----------------------------------------------------------------------------------------------------
\begin{figure}
  \centering
     \includegraphics[width=0.40\textwidth]{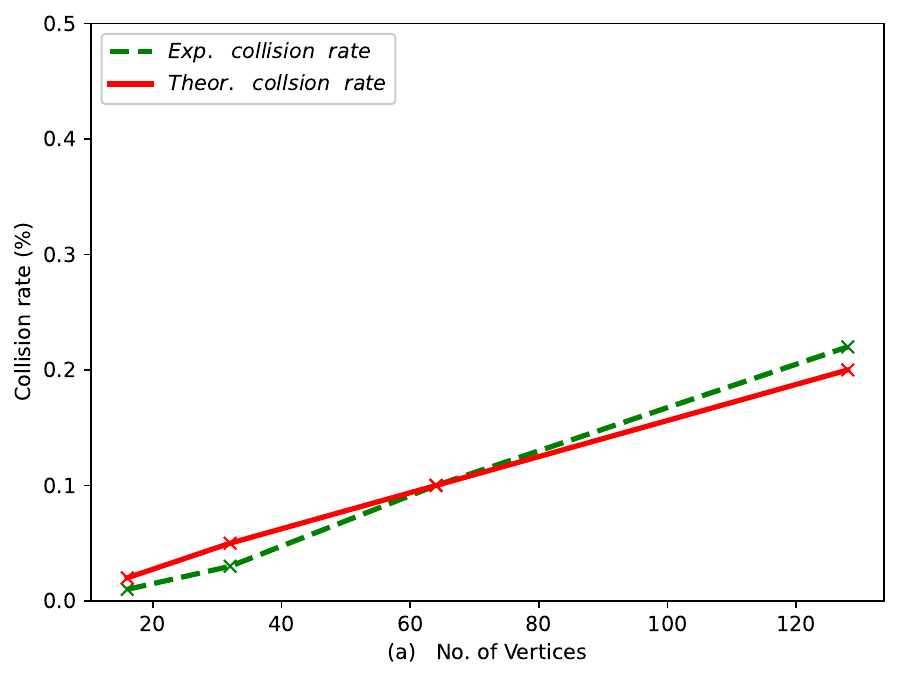}
          
       \caption{Experimental (green) and theoretical (red) collision rate  based on $N = 10000$ pair of randomly chosen  $1.5 N_v$-bit length messages  as function of the number of vertices of the HN4 network for $l=7, k=16$.}
 \label{collision_rate}      
\end{figure}
%----------------------------------------------------------------------------------------------------
%%%%%%%%%%%%%%%%%%%%%%%%%
\subsection{Uniform distribution} \label{dis}
%%%%%%%%%%%%%%%%%%%%%%%%%
In this sub-section we study the  distribution of  bits in the  hash values and look for the statistical signature of the original message in the hash values.  The following steps are performed: 
\begin{enumerate}
\item Perform $1. -3.$ from sub-section \ref{diff_hash}.

\item Count the  number of flipped bit at a location for  $N=10000$ pair of samples.  Let  $h(m_1) = \left(b_1^1, b_2^1, \cdots, b_L^1\right)$, $h(m_2) = \left(b_1^2, b_2^2, \cdots, b_L^2\right)$ are  the hash values in binary form.  Then the number of flipped bit-pairs, $T_i$,  at $i_{th}$ position  for $N$ experiments is given by 
$T_i = \sum_{j=0}^{N-1} \delta_j\left(b_i^1 - \bar{b_i^2}\right)$,  where $\bar{b_i^2}$ is   $NOT$  $b_i^2$.  $\delta_j(x)= 1$ for $x=0$,  and $\delta_j(x)= 0$ for $x \neq 0$. $T_i$ is plotted in fig.  \ref{hash_bar}(upper panel) as a function of the bit location. 

\item Repeat  $1. $  and   count the  number of  same bit-pairs,  $\tilde{T_i} = \sum_{j=0}^{N-1} \delta_j\left(b_i^1 - b_i^2 \right)$,   at a location for $N=10000$ pair of samples. 
$\tilde{T_i}$ is plotted in fig.  \ref{hash_bar}(lower panel) as a function of the bit location. 
\end{enumerate}
We can also estimate the mean of  flipped bit count  \cite{zhou} over the bit locations as 
\begin{eqnarray}
\bar{T} =  \sum_{i=1}^{L} T_i  /L \,,
 \label{mean}
\end{eqnarray}
and  the  standard deviation 
\begin{eqnarray}
\Delta T =   \sqrt{\sum_{i=1}^{L} \left(T_i - \bar{T} \right)^2 /(L-1)} \,. 
 \label{mean}
\end{eqnarray}
For $N=10000$ experiments, done in ideal conditions,  one would expect the mean $\bar{T}$ to be  $N/2 = 5000$.  The departure of the mean  from the ideal value  $|\bar{T}- N/2|$ along with $\bar{T}$ and $\Delta T$ are   presented in the table \ref{table3}.  

%----------------------------------------------------------------------------------------------------
\begin{figure}
  \centering
     \includegraphics[width=0.40\textwidth]{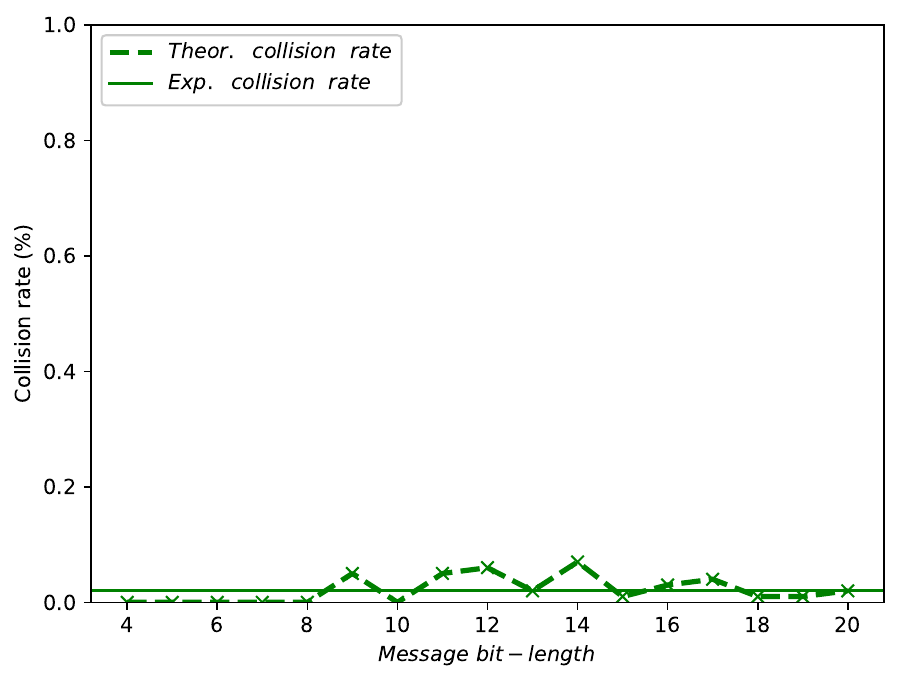}
          
       \caption{Experimental (solid green) and theoretical (dashed green) collision rates  as function of   $N = 10000$ pair of randomly chosen  messages  of varying bit-length  with  HN4 network for $l=6, k=16$.}
 \label{short_msg}      
\end{figure}
%----------------------------------------------------------------------------------------------------
%-------------------------------------------------------------------------------------------------------------------------------------------------------------
\begingroup

\setlength{\tabcolsep}{3.0pt} % Default value: 6pt
\renewcommand{\arraystretch}{1.2} % Default value: 1

\begin{table}
 \begin{center}

    \begin{tabular}{ |  p{3.0cm }  | p{1.2cm } |  p{1.6cm} | p{1.2cm} | }
    \hline
     $\textbf{ \scriptsize Hash function}$& $\bar{T}$ & $|\bar{T}-N/2|$  &  $\Delta T$   \\ \hline
    
    $256$-$\textbf{\scriptsize bit hash function}$ & 4790.92  & 209.08  &  10.78 \\ \hline
    
\end{tabular}
 \caption{Result for uniform distribution property}  
 \label{table3}
\end{center}
 \end{table} 
 \endgroup
 %-------------------------------------------------------------------------------------------------------------------------------------------------------------
 
 %-------------------------------------------------------------------------------------------------------------------------------------------------------------
\begingroup

\setlength{\tabcolsep}{3.0pt} % Default value: 6pt
\renewcommand{\arraystretch}{1.2} % Default value: 1

\begin{table}
 \begin{center}

    \begin{tabular}{ |  p{2.2cm }  | p{1.0cm } |  p{1.0cm} | p{1.0cm} | }
    \hline
     & $\omega = 0$ & $\omega = 1$  &  $\omega \geq 2$   \\ \hline
    
    $W^E\left(\omega \right)$ & $9995$ & $5$  & $0$  \\ \hline
    
    $W^T \left(\omega \right)$  & $9997$ & $2$  & $0$  \\ \hline

\end{tabular}
 \caption{Experimental and theoretical values of collision.} 
  \label{table4}
\end{center}
 \end{table} 
 \endgroup
 %-------------------------------------------------------------------------------------------------------------------------------------------------------------

%-------------------------------------------------------------------------------------------------------------------------------------------------------------
\begingroup

\setlength{\tabcolsep}{3.0pt} % Default value: 6pt
\renewcommand{\arraystretch}{1.2} % Default value: 1
\begin{table*}
\begin{center}
\begin{tabular}{|  p{1.6cm } |  p{1.6cm }  | p{1.6cm } |  p{1.6cm} | p{1.6cm} | p{1.6cm} | p{1.65cm} |}
\hline
     
   & Ref. \cite{li} & Ref. \cite{dan} & Ref. \cite{yang} & Ref. \cite{cao} &  Ref. \cite{yang1} & Our scheme \\ \hline
    
Exp.   & $23.12$  \cite{yang1} & $10.18$ & $6.33$ & $1.95$ & $1.46$ &  $0.05$ \\ \hline
    
Theor.   &  $9.32$  & $9.32$ & $0.06$ & $0.30$ & $0.21$ &  $0.02$ \\ \hline

\end{tabular}
 \caption{Collision rate(\%) of different hash function schemes.} 
 \label{table5}
\end{center}
 \end{table*} 
 \endgroup
 %-------------------------------------------------------------------------------------------------------------------------------------------------------------
 
 %-------------------------------------------------------------------------------------------------------------------------------------------------------------
\begingroup

\setlength{\tabcolsep}{3.0pt} % Default value: 6pt
\renewcommand{\arraystretch}{1.2} % Default value: 1
\begin{table*}
\begin{center}
\begin{tabular}{|  p{1.65cm } |  p{1.6cm } |  p{5.5cm } | p{7.0cm } | }
\hline
     
   & Coin dim.  & Initial state & Evolution operators  \\ \hline
    
Ref. \cite{li}   & $4$  & $|0 0\rangle_v \otimes (\alpha |00\rangle + \beta |01\rangle + \chi |10\rangle + \delta |11\rangle)_c$  &$S_{xy}(I\otimes C_0)/ S_{xy}(I\otimes C_1)$  \\ \hline
    
Ref. \cite{dan}   &  $2$  & $|0\rangle_v \otimes (\alpha |0\rangle + \beta |1\rangle)_c$  & $S_y(I\otimes C_0)S_x(I\otimes C_0) / S_y(I\otimes C_1)S_x(I\otimes C_1)$  \\ \hline

Ref. \cite{yang}   & $4$ & $|0\rangle_v \otimes (\alpha |00\rangle + \beta |01\rangle + \chi |10\rangle + \delta |11\rangle)_c$   & $S_{xy}(I\otimes C_0)/ S_{xy}(I\otimes C_1)$  \\ \hline
    
Ref. \cite{cao}   &  $n-1$  & $|0\rangle_v \otimes  \frac{1}{\sqrt{n-1}} \sum_{i=1}^{n-1} |i\rangle_c$  & $S_J(I\otimes C_{0})/ S_J(I\otimes C_{1})$  \\ \hline

Ref. \cite{yang1}   & $2$   & $|0\rangle_v \otimes (\cos\alpha |0\rangle + \sin\alpha |1\rangle)_c$  &$S_c(I\otimes C_{00})/ S_c(I\otimes C_{01})/ S_c(I\otimes C_{10})/ S_c(I\otimes C_{11})$ \\ \hline
    
Our scheme   &  $4$  & $\frac{1}{2} \sum_{i=0}^3 |i\rangle_c \otimes |0\rangle_v$ & $S_0(C_0\otimes I)/ S_1(C_1\otimes I)$  \\ \hline

\end{tabular}
 \caption{Comparison of resources  of different hash function schemes.} 
 \label{table6}
\end{center}
 \end{table*} 
 \endgroup
 %-------------------------------------------------------------------------------------------------------------------------------------------------------------
 %----------------------------------------------------------------------------------------------------
\begin{figure}
  \centering
     \includegraphics[width=0.40\textwidth]{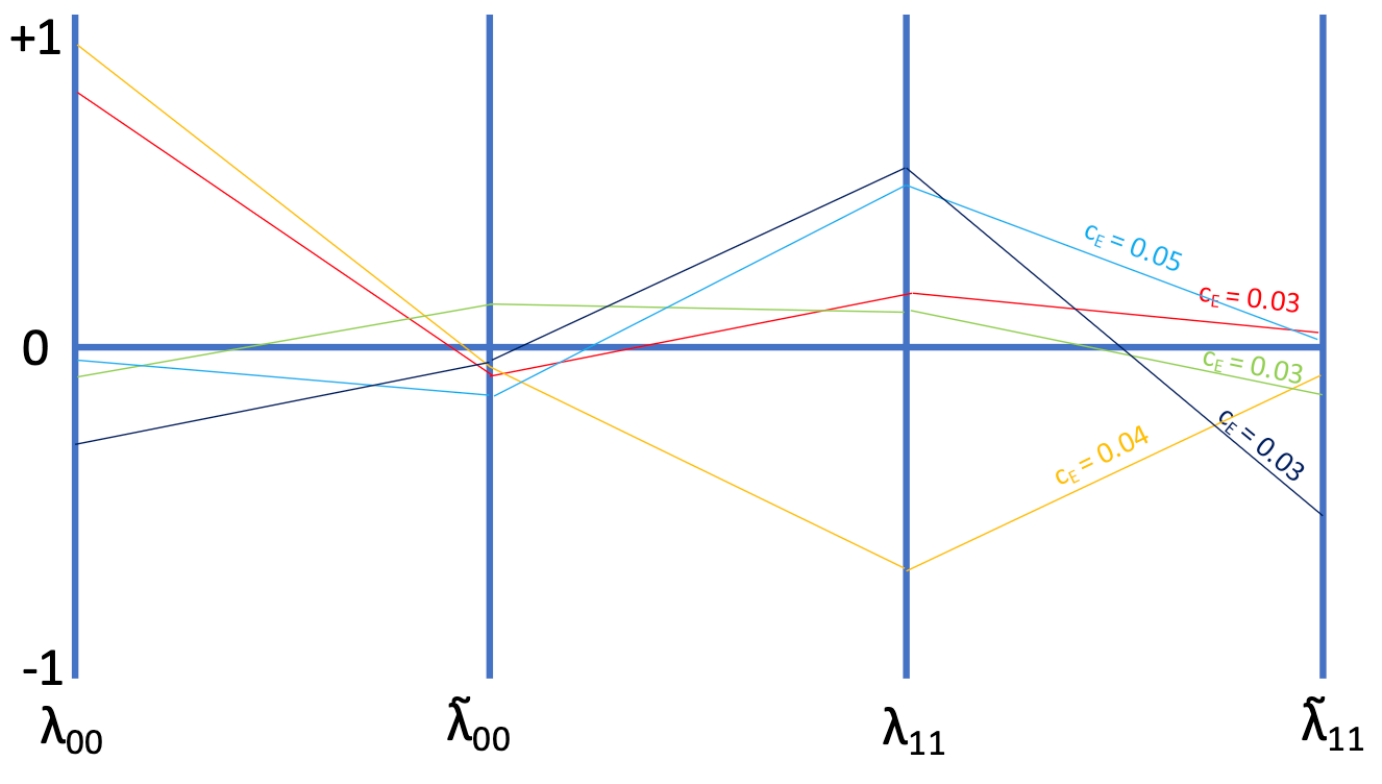}
          
       \caption{Comparison of collision rates for five randomly chosen set of long-range weights  $\lambda_{00}, \tilde{\lambda}_{00}, \lambda_{11}, \tilde{\lambda}_{11},$ of the HN4 network.}
 \label{fig7}      
\end{figure}
%----------------------------------------------------------------------------------------------------
%%%%%%%%%%%%%%%%%%%%%%%%%
\subsection{Collision analysis} \label{coll}
%%%%%%%%%%%%%%%%%%%%%%%%%
One of the most  important things to keep in mind, while designing a hash function, is that  different messages are not mapped  to the same hash value.  This property is known as collision resistance, which is hard to  prove for  a   hash function. However, we can study collision resistance property  of our  hash   scheme by    estimating  the number of times $\omega$  there is a match  between the two  numbers  at  the same location  of   a pair of hash values corresponding to two different messages.  The following experiment is performed:

\begin{enumerate}
\item Perform $1. -2.$ from subsection \ref{diff_hash}. 

\item Generate $256$-bit  hash values, $h(m_1), h(m_2)$, in  decimal format,   corresponding  to the pair of messages $m_1, m_2$ respectively.  For  the   quantum walk with $N_v$ vertices, there are $N_v$ decimal numbers in a  hash value.

\item For the pair of hash values   $h(m_1)$ and $h(m_2)$,  count the  number of pairs, which have   same decimal numbers at same locations.  
Let  $h(m_1) = \left(d_1^1, d_2^1, \cdots, d_{N_v}^1\right)$, $h(m_2) = \left(d_1^2, d_2^2, \cdots, d_{N_v}^2\right)$ are  the hash values in decimal  form.  Then the number of same decimal pairs  at same location, $ \omega = \sum_{i=1}^{N_v} \delta\left(d_i^1 - d_i^2 \right)$,   $\omega = 0, 1, 2, \cdots, N_v$.

\item   Repeat  $1. - 3.$  for $N=10000$ times to obtain  the number of pairs $W^E(\omega)$ having  a fixed $\omega$. 
\end{enumerate}
Theoretical   value can be calculated assuming  binomial distribution as 
\begin{eqnarray} \nonumber
W^T(\omega) =&& N \times \scriptsize{prob}(\omega) = \\
&&N \frac{N_v!}{\omega ! (N_v- \omega)!} \left( \frac{1}{2^{k}}\right)^{\omega} \left( 1- \frac{1}{2^{k}}\right)^{N_v-\omega}\,,
 \label{mean}
\end{eqnarray}
$W^T(\omega)$ and $W^E(\omega)$  presented in table \ref{table4}  shows that the experimental values are close to the theoretical values. The experimental collision rare of   $c_E =W^E(\omega)\times 100=0.05\%$  is very close to the theoretical collision rate of  $c_T = W^T(\omega)\times 100=0.02\%$,   indicating  very  strong   collision resistance.  
Note that such high collision resistance  of our hash function  is attributed to the suitable  choice of  values  for  the long-range-edge  weights  $l_{00}, \tilde{l}_{00}, l_{11}, \tilde{l}_{11}$. The set of values  $l_{00} = 0.01, \tilde{l}_{00} = 1.0,  l_{11} = 0.1,  \tilde{l}_{11} =0.01$ chosen heuristically  in this  article is one of  the many such  suitable set of values for which  the experimental collision rate  $c_E$ is close to the theoretical collision rate  $c_T = 0.02\%$.  
The only requirement  is that these parameters cannot be zero,  since  for zero long-range weights HN4 network reduces to  a cycle, which has $100 \%$ collision rate.  
In Fig. \ref{fig7} we  evaluated collision rates for five randomly chosen sets  of  values  of the normalized long-range weights  $( \lambda_{00} = \mathcal{N}_0\sqrt{l_{00}},  \tilde{\lambda}_{00} = \mathcal{N}_0\sqrt{\tilde{l}_{00}}, \lambda_{11} = \mathcal{N}_1\sqrt{l_{11}}, \tilde{\lambda}_{11} = \mathcal{N}_1\sqrt{\tilde{l}_{11}})$   and  found that the collision rates are very close to the theoretical value  $c_T = 0.02$ despite wide variations in the parameters, implying any random choices of the long-range weights  are    suitable  for the  hash function. 
We also evaluated trace distance  $D_{m1,m2} = \sqrt{1- |\langle \psi_{m1}| \psi_{m2} \rangle|^2}$  between the quantum output states $|\psi_{m1}\rangle, |\psi_{m2}\rangle$ corresponding to $N=10000$ randomly chosen  pair of messages $(m1, m2)$ and found that trace distance satisfies  $D_{m1,m2} \geq 1-\epsilon$ with $\epsilon \leq 0.25$, suggesting strong separability  between  the  output quantum states. 
%%%%%%%%%%%%%%%%%%%%%%%%%
\subsection{Resistance to birthday attack} \label{birth}
%%%%%%%%%%%%%%%%%%%%%%%%%
In this paper we have discussed hash values with a bit-length of $L= N_v \times k = 16\times 16 = 256$, which  means  we need  $2^{L/2} = 2^{128}$ trials to find if   two hash values  collide with a probability of $0.5$, which suggests high resistance to birthday attack.  Moreover, we can further increase bit-length of the hash value by increasing the length  of  the lattice $N_v$.  
In cryptography,  fixed-size  classical hash functions up-to $512$-bit hash values are currently  being used in practice.  Our hash function can easily be extended to $512$-bit hash value,  corresponding  to  $N_v = 32$, and beyond  with  high collision resistance property by appropriately tuning  the precision level as can be seen from Fig. \ref{collision_rate}.   It   has good scaling property as the  collision rate of  the  hash function, represented by the dashed green curve is very   close to  the theoretical collision rate, represented by the solid red curve.   The parameters $l_{00}, \tilde{l}_{00}, l_{11}, \tilde{l}_{11}$  remain same as the previous experiments in the analysis of  Fig. \ref{collision_rate}, while  the numbers of vertices  $N_v = 8, 16, 32, 64, 128, 256, 512$ are considered  for the evaluation. 
%----------------------------------------------------------------------------------------------------
\begin{figure}
  \centering
     \includegraphics[width=0.40\textwidth]{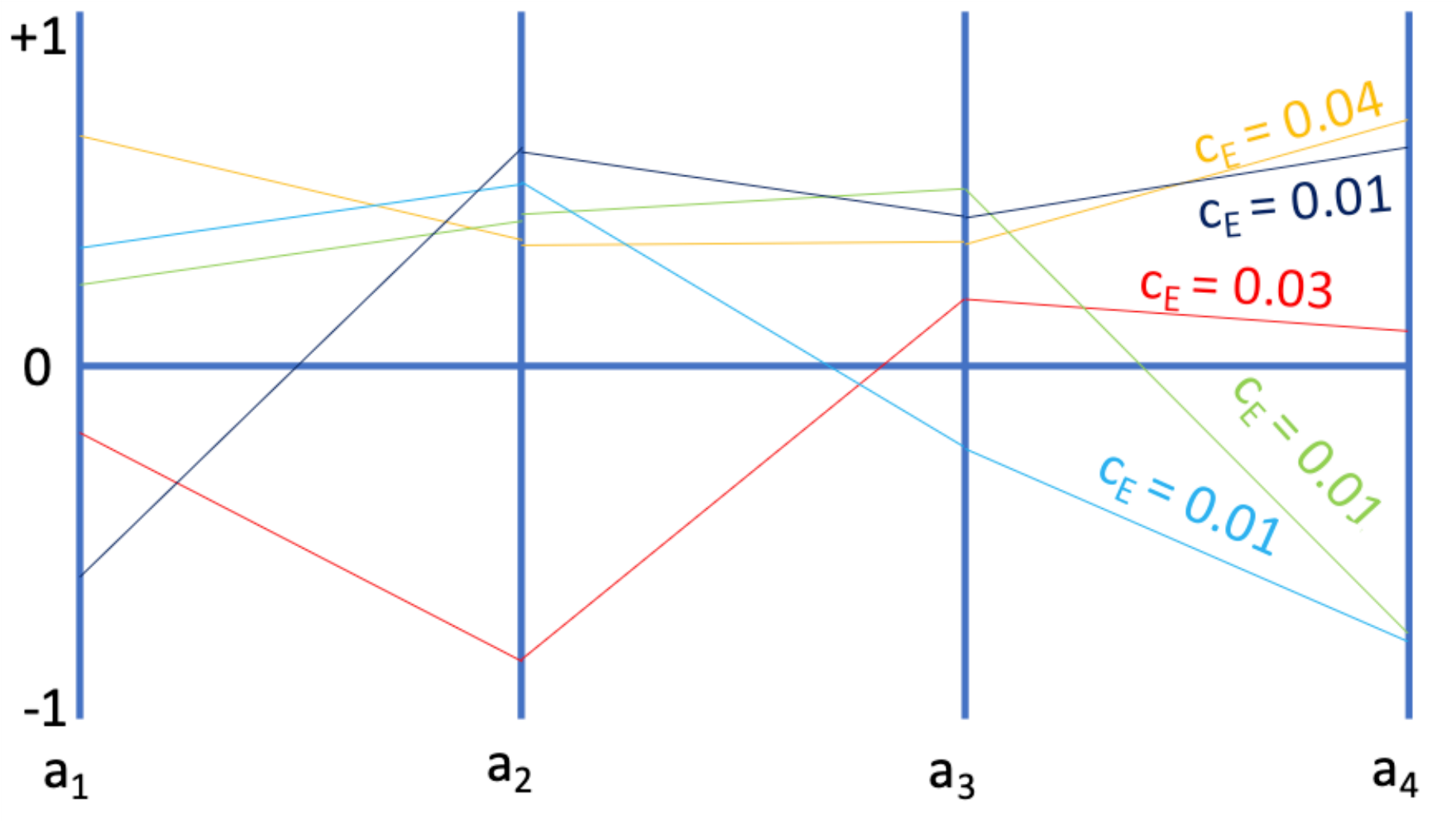}
          
       \caption{Comparison of collision rates for five randomly chosen set of amplitudes  $a_1, a_2, a_3, a_4$ of the  initial coin state.}
 \label{fig8}      
\end{figure}
%----------------------------------------------------------------------------------------------------
%%%%%%%%%%%%%%%%%%%%%%%%%
\section{Comparison with other quantum walk based  hash schemes} \label{comp}
%%%%%%%%%%%%%%%%%%%%%%%%%
One of the most important indicators  for a  secured hash function is the collision rate.  The lower the collision rate the better is the hash function.  A  comparison of  the collision rates(experimental and theoretical) of different  hash  schemes is   presented in  table \ref{table5}, where the theoretical values are obtained using the binomial distribution in  eq. (\ref{mean}).  Since the required details for calculating  $W^T$  for ref. \cite{li}  is not available,  we calculate  its  theoretical value of collision  in  table  \ref{table5},  assuming  $N_v= 25$ vertices and each vertex contributing  $8$ bit  to a  $200$-bit hash value and performing collision test for $N=10000$ times.  Collision rate of our hash scheme,  presented in the   seventh column of table  \ref{table5}, shows that the experimental value $0.05$ is very close to the theoretical value $0.02$.  Collision rate of our scheme can further be improved using higher precision label as shown in fig.  \ref{collision_rate}.   

A comparison of   the resource requirements of our scheme  along with others are presented in table \ref{table6}.   

Another advantage of our scheme is that, it   works well  for short messages also,   as  demonstrated in  Fig. \ref{short_msg}.  The experimental collision rate, represented by the dashed green curve  for $N_v=16$  and message bit-length $4-20$, remains  close to the theoretical collision rate represented by the solid green line.  Although, for very short message bit-length $< 4$, the  collision rate increases significantly, because the probability does not have enough time to  spread.  On the other hand since  there is predictable zero probability amplitudes for  quantum walk  on even-length one-dimensional lattice as already discussed  in Section \ref{hanoi}, which results in $100 \%$ collision rate for the hash function. This is the reason most of the quantum walk based hash schemes based on one-dimensional periodic lattices   cannot use  even-length  lattice at all. Also they cannot use message bit-length $< N_v$  for odd-length one-dimensional periodic lattice as the collision rate is very high.

Although the initial state for the analysis is chosen to be equal superposition of all the coin basis states at the origin of the vertex, i.e.,  $|\psi_{in}\rangle = \frac{1}{2} \sum_{i=0}^3 |i\rangle \otimes |x_v=0\rangle$ for simplicity, however, we can randomly choose the amplitudes  $a_1, a_2, a_3, a_4$ of the initial state $|\psi_{in}\rangle = \left( a_1|0\rangle  + a_2|1\rangle+ a_3|2\rangle+ a_4|3\rangle\right)\otimes |x_v=0\rangle$ to carry out our analysis without  compromising performance. In Fig. \ref{fig8} we  evaluated collision rates for five randomly chosen sets  of amplitudes and  found that the collision rates are very close to the theoretical value  $c_T = 0.02$ despite wide variations in the amplitudes, implying any random choices of the initial state are  suitable for the  hash function.

Two  important  aspects of our proposed scheme, which are responsible for such high collision resistance property,  are the following:   Firstly, we have introduced two additional long-range edges  to  the  two regular edges that  allow  us to tune the parameters in such a way that  the probability distribution can be  modified.  Secondly, in addition to the coin operator we also control the shift operator by the message bits as opposed to the other  quantum walk based hash functions which only control  the coin operator by the message.

%%%%%%%%%%%%%%%%%%%%%%%%%
\section{Conclusion} \label{con}
%%%%%%%%%%%%%%%%%%%%%%%%%
Quantum  walk can be  used to generate hash function by controlling the evolution operator   $ \mathcal{U} = S \left(C \otimes I \right)$.  In most of the  works in the literature, message    controls only the coin operator $C$, depending on the bit values  of the  message. However, in this paper we are controlling both the coin operator  $C$ and shift operator $S$.  For the bit value $0$, we use the  evolution operator $ \mathcal{U}_0 = S_0 \left(C_0 \otimes I \right)$ and for the bit value $1$, we use the evolution operator $ \mathcal{U}_1 = S_1 \left(C_1 \otimes I \right)$, where 
$C_0, C_1$ are Grover operators with different underlying coin states controlled by the bit values    and $S_0, S_1$ are   flip-flop and standard shift operator  respectively  for the HN4 network.

Our proposed  quantum walk based hash function is executed on the one-dimensional periodic lattice, equipped with extra long-range edges,  known as   HN4 network.    It allows us to obtain a hash function, which is  highly resistant to the collision  in comparison with other quantum walk based hash functions  as can be seen from table \ref{table5} that  the collision rate  is very low.  
It is also  important to note  that the the security of our scheme,  including  any other quantum hash schemes,    is guaranteed by the Holevo theorem \cite{holevo}, whereas, security of classical hash functions is  based on computational security of a hard  mathematical  function, that can be challenged by quantum  algorithms.

The effect of noise on the security of hash function  is not considered in  this article.   However, applications  based on quantum walk, and more generally   quantum computing models are  known  to be susceptible to  noise.  Specifically, in our case,   unitary noise  is  important,   as longer  messages  require  more  iteration steps of the quantum walk,  contributing to greater   decoherence related effect   in the final state. 
Note that the noise usually disrupts the precision of the quantum operations, which can affect the deterministic nature of quantum hash function and ultimately  affecting security  of a quantum-based hash function as suggested in ref. \cite{upa}. In this context it also should be noted that  not all noises are bad, as reported in ref.  \cite{yyang},  broken-line  decoherence type noise   on a  cycle   improves the performance of the quantum walk based hash function. Detailed investigation on the pros and cons of different types of noises  on the quantum-based hash function,  including  quantum walk based hash function can be an interesting future research direction. 
One of the possible  noise mitigation techniques  in our scheme  would be to  reduce the required  number of iteration steps for a given message length    by allowing more than one bit of the message to control  the evolution operator at a  fixed  iteration  step. It  will also  be interesting  to see  how   the  long-range edges  of the HN4 network  affect the  unwanted   noise.

\vspace{1cm}

Data availability Statement:  The data  generated during and/or analyzed during the current study is  included in the article.
\vspace{0.5cm}

Conflict of interest: The authors have no competing interests to declare that are relevant to the content of this article.

%%%%%%%%%%%%%%%%%%%%%%%%%%%

\end{document}